\documentclass[aps,prb,twocolumn,epsf,floatfix]{revtex4}
\usepackage{graphicx}
\usepackage{amsmath}
\begin{document}
\title{Spin Relaxation in a Two-Dimensional Electron Gas in a Perpendicular
Magnetic Field}
\author{A.A. Burkov and Leon Balents}
\affiliation{\small Department of Physics, University of California,
Santa Barbara, CA 93106} 
\date{\today}

\begin{abstract}
We consider the problem of spin relaxation in a two-dimensional electron gas
(2DEG) in a perpendicular magnetic field.
We assume that the spin relaxation is induced by the Rashba spin-orbit (SO)
interaction, which appears due to the inversion asymmetry of the confining
potential.   
Our solution is based on a microscopic evaluation of the spin density response
function of the 2DEG with impurities and SO interaction.
We derive explicit expressions for the transverse and longitudinal 
spin relaxation rates. 
Our analysis shows, in particular, that the spin relaxation rates exhibit 
{\it magnetoquantum oscillations}, which are analogous to the 
Shubnikov-de Haas oscillations of the electrical resistivity.
These oscillations can be observed, for example, in time-resolved optical 
experiments.
    
\end{abstract}

\maketitle
\section{Introduction}
\label{sec:1}
The study of spin relaxation in semiconductor 
2DEGs is an important area of the emerging field of {\it spintronics}.\cite{Awschalom02}
The dominant spin relaxation mechanism in these
systems can typically be associated \cite{Dyakonov86} with the Rashba SO 
interaction,\cite{Rashba84} which exists due to the inversion asymmetry of 
the confining potential in semiconductor heterostructure based 2DEG systems.
A simple semiclassical picture of the spin relaxation, due to Dyakonov and 
Perel (DP),\cite{Dyakonov71,Dyakonov86} is that since the Rashba 
interaction has the form of a momentum-dependent magnetic 
field $\lambda \hat z 
\cdot [\boldsymbol{\tau} \times {\bf p}]$, it induces precession of the spin 
of a moving electron. 
This precession leads to the randomization of the spin at long times.    
The DP spin relaxation rate is
given, to leading order in $\lambda$, by the simple expression 
$1/\tau_{sf} \sim (\lambda p_F/\hbar)^2 \tau$, 
where $p_F$ is the Fermi momentum and $\tau$ is the elastic scattering time. 
Quasiclassically, $\lambda p_F/\hbar$ is simply the spin-precession frequency,
associated with the Rashba field. 
The distinguishing feature of the DP spin relaxation is that the relaxation 
rate is smaller in more disordered systems, 
since impurity scattering disrupts the Rashba spin precession by randomizing 
the electron's momentum. 

It is well-known that transport processes in 2DEG systems are strongly 
influenced by the application of a perpendicular magnetic field. 
The perpendicular field quantizes electron's energy spectrum, which 
manifests spectacularly in Shubnikov-de Haas (SdH) oscillations of the 
resistivity and eventually leads to the quantum Hall effect.
In this paper we address the question of how a perpendicular 
magnetic field influences the Rashba SO interaction-induced spin relaxation 
in semiconductor 2DEG systems.
This problem has been addressed before by several 
authors, both in the 
semiclassical \cite{Ivchenko73,Margulis83,Martin02,Puller03} and in the 
quantum limits,\cite{Frenkel91,Dikman96,Bychkov99,Nazarov00,Khaetskii01,Dikman03} but a complete and rigorous analysis is still lacking. 

In this paper we will provide such an analysis. 
We will mainly be interested in the regime of moderate magnetic fields, such
that $\omega_c \tau \lesssim 1$ (here $\omega_c$ is the cyclotron frequency), 
and $\epsilon_F \gg \hbar \omega_c$. 
Our analysis can, however, be generalized to the quantum Hall limit, when
$\epsilon_F \sim \hbar \omega_c$ and $\omega_c \tau$ is large.
Our theory is based on a 
microscopic evaluation of the spin density response function of a disordered
2DEG system with the Rashba SO interaction in a perpendicular magnetic field. 
We obtain explicit analytical expressions for the longitudinal and 
transverse spin relaxation rates:
\begin{eqnarray}
\frac{1}{\tau_z} &=& \frac{8 \lambda^2 m \epsilon_F \tau /\hbar^2}
{1 + (\omega_c \tau)^2}, \nonumber \\
\frac{1}{\tau_{\perp}} &=& \frac{1}{2\tau_z},
\nonumber
\end{eqnarray}
which reduce to the well-known Dyakonov-Kachorovskii
\cite{Dyakonov86} expressions in the zero-field limit.
We show, in particular, that the application of a perpendicular magnetic 
field leads to {\it quantum 
oscillations} of the spin relaxation rate, which are analogous to the 
well-known quantum oscillations of transport coefficients.
These oscillations arise from the magnetic field dependence of the 
scattering time $\tau$, which will be calculated below. 

The paper is organized as follows.
In section \ref{sec:2} we calculate the disorder-averaged Green's
function of a 2DEG system with the Rashba SO interaction in a perpendicular 
magnetic field using the self-consistent Born approximation (SCBA). 
In section \ref{sec:3} the spin-density response function of our system
is calculated by summing SCBA self-energy and ladder vertex corrections to the
bare polarization diagram. 
The transverse and longitudinal spin relaxation rates 
are then extracted from the poles of this response function.
\section{Density of states and disorder self-energy}
\label{sec:2}
The simplest single-particle Hamiltonian describing the dynamics of 
electrons in a semiconductor 2DEG in a perpendicular magnetic field 
can be written as:
\begin{equation}
\label{eq:1}
H_0=\frac{\boldsymbol{\pi}^2}{2m} + \lambda \hat z \cdot [\boldsymbol{\tau}
\times \boldsymbol{\pi}] - \frac{\Delta_z}{2} \tau^z,
\end{equation}
where $\boldsymbol{\pi}={\bf p}+(e/c){\bf A}$ is the kinetic momentum,
$\boldsymbol{\tau}$ is the spin operator and $\Delta_z=g\mu_B B$ is the Zeeman 
energy. 
The second term in Eq.(\ref{eq:1}) describes the Rashba SO interaction. 
Hamiltonian $H_0$ can be easily diagonalized \cite{Das90,Falko92,Schliemann03}
if one notices that the Rashba term mixes only neighboring Landau levels
with opposite spin directions.
The eigenstates are thus given by: 
\begin{equation}
\label{eq:2}
|n,a\rangle = u_{na} |n,\downarrow \rangle + v_{na} |n-1,\uparrow \rangle,
\end{equation}
where $a=1,2$ and $|n,\sigma \rangle$ is the $n$-th Landau level eigenstate
with spin $\sigma=\uparrow,\downarrow$.
The corresponding eigen-energies are:
\begin{equation}
\label{eq:3}
\epsilon_{na} = \hbar \omega_c n + (-1)^a \sqrt{\left(\frac{\hbar \omega_c +
\Delta_z}{2}\right)^2 + 2 \lambda^2 m \hbar \omega_c n},
\end{equation}
where $\omega_c = eB/mc$.
Note that $n=1,2,...$ for $a=1$, but $n=0,1,...$ for $a=2$. 
The amplitudes $u_{na}$ and $v_{na}$ are given by:
\begin{eqnarray}
\label{eq:4}
u_{na}&=&i \sqrt{\frac{1}{2} + \frac{(-1)^a(\hbar \omega_c + \Delta_z)/2}
{\sqrt{(\hbar \omega_c + \Delta_z)^2/4 + 2 \lambda^2 m \hbar \omega_c n}}},
\nonumber \\
v_{na}&=&(-1)^{a+1}\sqrt{\frac{1}{2} +  
\frac{(-1)^{a+1}(\hbar \omega_c + \Delta_z)/2}
{\sqrt{(\hbar \omega_c + \Delta_z)^2/4 + 2 \lambda^2 m \hbar \omega_c n}}}. 
\nonumber \\
\end{eqnarray}

Using the above basis of single-particle eigenstates, the Hamiltonian
of the 2DEG system in the presence of impurity scattering potential can be 
written as:
\begin{equation}
\label{eq:5}
H=\sum_{nak}\epsilon_{na} c^{\dag}_{nak} c^{\vphantom\dag}_{nak} +
\sum_{nak,n'a'k'} V_{na,n'a'}(k,k') c^{\dag}_{nak} c^{\vphantom\dag}_{n'a'k'},
\end{equation}
where index $k$ denotes orbit-center quantum numbers in the Landau gauge and
\begin{equation}
\label{eq:6}
V_{na,n'a'}(k,k')=\sum_{\sigma} \int d{\bf r} \Psi^*_{nak} ({\bf r}\sigma)
V({\bf r}) \Psi^{\vphantom *}_{n'a'k'}({\bf r} \sigma),
\end{equation}
are the matrix elements of the impurity potential $V({\bf r})$ in the basis 
(\ref{eq:2}).
The wavefunctions $\Psi_{nak}({\bf r}\sigma)$ are given by:
\begin{eqnarray}
\label{eq:7}
\Psi_{nak}({\bf r} \uparrow)&=&v_{na} \phi_{n-1,k}({\bf r}), \nonumber \\
\Psi_{nak}({\bf r} \downarrow)&=&u_{na} \phi_{n,k}({\bf r}),
\end{eqnarray}
where $\phi_{n,k}({\bf r})$ are the Landau-level eigenfunctions in the Landau
gauge.

Our analysis of the spin relaxation in a system, described by Eq.(\ref{eq:5}),
is based on a calculation of the spin density response function, which 
is similar to Ando's calculation of the conductivity of a 2DEG in a 
perpendicular field.\cite{Ando} 
Our calculation is technically more complicated than Ando's due to the 
presence of the SO interactions and also due to the fact that vertex 
corrections to the polarization bubble,
which vanish in the case of the conductivity calculation, are crucial
in our case. 

Following Ando, we use the self-consistent Born approximation (SCBA) 
to find the disorder-averaged Green's function.
This is a very good approximation for the high Landau level filling factors
that we are assuming.
SCBA equation for the retarded disorder self-energy in our case reads:
\begin{equation}
\label{eq:8}
\Sigma_{na}^R(\epsilon) = \sum_{n'a'} \Gamma_{na,n'a'} 
G^R_{n'a'}(\epsilon),
\end{equation}
where 
\begin{equation}
\label{eq:9}
\Gamma_{na,n'a'} = \sum_{k'} \langle V_{na,n'a'}(k,k') V_{n'a',na}(k',k)
\rangle,
\end{equation}
and
\begin{equation}
\label{eq:10}
G^R_{na}(\epsilon) = \frac{1}{\epsilon-\epsilon_{na}-
\Sigma_{na}^R(\epsilon)},
\end{equation}
is the retarded SCBA Green's function. 
Angular brackets in Eq.(\ref{eq:9}) denote disorder average.

Matrix elements of $\Gamma$ can be easily evaluated for $\delta$-function
impurity potential and are given by:
\begin{equation}
\label{eq:11}
\Gamma_{na,n'a'}=\frac{\hbar^2 \omega_c}{2\pi\tau_0}\left(|u_{na}|^2
|u_{n'a'}|^2 + |v_{na}|^2 |v_{n'a'}|^2\right),
\end{equation}
where $\tau_0$ is the elastic impurity scattering time in the absence of 
the magnetic field.

Eq.(\ref{eq:8}) is too complicated to be solved analytically without 
approximations. 
Fortunately, only very minor simplification is needed to make this
equation solvable.
The main complication is the dependence of the matrix $\Gamma$ on the
Landau level index $n$. 
This dependence is, however, inessential, since it appears entirely 
due to the term $2\lambda^2 m\hbar \omega_c n$ in Eq.(\ref{eq:4}). 
The dependence on $n$ in this term can be safely ignored, since 
the Rashba SO interaction can be assumed to be a weak perturbation, in the 
sense that $\lambda p_F \ll \epsilon_F$.     
In this case we can make a replacement 
$n \rightarrow \epsilon_F/\hbar \omega_c$ 
in the Rashba term in Eqs.(\ref{eq:3}) and (\ref{eq:4}), which makes the 
amplitudes $u_{na}$ and $v_{na}$ and, consequently, $\Gamma$, independent of
$n$.
This simple approximation allows us to solve Eq.(\ref{eq:8}) 
analytically. 

It is convenient to introduce the following new notation.
Let
\begin{equation}
\label{eq:12} 
\Delta = 2 \sqrt{\left(\frac{\hbar \omega_c +\Delta_z}{2} \right)^2 + 
2\lambda^2 m \epsilon_F}.
\end{equation}
Then $\epsilon_{na} = \hbar \omega_c n + (-1)^a \Delta/2$.
Also define:
\begin{eqnarray}
\label{eq:13}
&&u_1 = i \cos(\vartheta), \,\, v_1 = \sin(\vartheta), \nonumber \\
&&u_2 = i \sin(\vartheta), \,\, v_2 = - \cos(\vartheta),
\end{eqnarray}
where
\begin{equation}
\label{eq:14}
\cos(\vartheta) = \sqrt{\frac{1}{2} - \frac{\hbar \omega_c + \Delta_z}
{2\Delta}}.
\end{equation}
Using this notation, the matrix elements of $\Gamma$ are given by:
\begin{eqnarray}
\label{eq:15}
\Gamma_{11} &=& \Gamma_{22} = \frac{\hbar^2 \omega_c}{2 \pi \tau_0} 
\left(1 - \frac{1}{2} \sin^2(2\vartheta) \right), \nonumber \\
\Gamma_{12} &=& \Gamma_{21} = \frac{\hbar^2 \omega_c}{\pi \tau_0} 
\sin^2(2\vartheta).
\end{eqnarray}
Let us also make the usual assumption that the real part of the disorder 
self-energy can be absorbed into the chemical potential.
Then SCBA equation can be written as:
\begin{equation}
\label{eq:16}
\textnormal{Im} \Sigma^R_a(\epsilon) = \sum_{n,a'} \Gamma_{aa'} 
\textnormal{Im} G^R_{na'} (\epsilon) =     
-2 \pi^2 \ell^2 \sum_{a'} \Gamma_{aa'} \varrho_{a'}(\epsilon),
\end{equation}
where 
\begin{equation}
\label{eq:17}
\varrho_a(\epsilon) = - \frac{1}{2\pi^2 \ell^2} \sum_n \textnormal{Im} 
G^R_{na}(\epsilon),
\end{equation}
is the density of states and $\ell=\sqrt{\hbar c/eB}$ is the magnetic length. 
Writing out Eq.(\ref{eq:17}) explicitly, we have:
\begin{eqnarray}
\label{eq:18}
&&\varrho_a(\epsilon) = - \frac{1}{2\pi^2 \ell^2} \nonumber \\
&\times& \sum_{n=0}^{\infty} 
\frac{\textnormal{Im} \Sigma^R_a(\epsilon)}
{\left[\epsilon - \hbar \omega_c n - (-1)^a \Delta - \hbar \omega_c 
\delta_{a,1}\right]^2 + \left[\textnormal{Im} \Sigma^R_{a}(\epsilon)\right]^2}.
\nonumber \\
\end{eqnarray}
The sum over Landau level indices in the above equation can be done
using Poisson summation formula: \cite{Golub01}
\begin{equation}
\label{eq:19}
\sum_{n=0}^{\infty} f(n) = \frac{f(0)}{2} + \sum_{p=-\infty}^{\infty}
\int_0^{\infty} dn f(n) e^{2\pi i p n}.
\end{equation}
Up to this point, our calculations were applicable to magnetic fields of 
general strength. 
To proceed further, we will restrict ourselves to the regime when 
$\epsilon_F \tau /\hbar \gg 1$ and $\epsilon_F \gg \hbar \omega_c$.
(The calculation we present can be done in other regimes as well, but the 
actual procedure will be a little different.) 
This allows us to neglect the term $f(0)/2$ in Eq.(\ref{eq:19}).
Then, changing the integration variable to $x=\hbar \omega_c n - \epsilon
+(-1)^a \Delta-\hbar \omega_c \delta_{a,1}$ and extending the lower limit 
of integration to $-\infty$, which is justified by the above assumptions, we
obtain:
\begin{eqnarray}
\label{eq:20}
&&\varrho_a(\epsilon) = -\frac{1}{2\pi^2\ell^2}  
\sum_{p=-\infty}^{\infty}
\int_{-\infty}^{\infty}\frac{dx}{\hbar \omega_c} \nonumber \\ 
&\times& \exp\left(2\pi i p
\frac{x+\epsilon-(-1)^a\Delta-\hbar\omega_c\delta_{a,1}}{\hbar\omega_c}\right)
\frac{\textnormal{Im} \Sigma^R_a(\epsilon)}{x^2 + 
\left[\Sigma^R_a(\epsilon)\right]^2}. \nonumber \\
\end{eqnarray} 
Calculating the above integral, we obtain:
\begin{eqnarray}
\label{eq:21}
\varrho_a(\epsilon) &=& \frac{m}{2\pi \hbar^2}\left[1 + 
2 \sum_{p=1}^{\infty}e^{-\pi p/\omega_c \tau_0} \right. \nonumber \\
&\times&\left.\cos\left(2\pi p \frac{\epsilon - (-1)^a\Delta -
\hbar\omega_c\delta_{a,1}}
{\hbar \omega_c}\right) \right],\nonumber \\
\end{eqnarray}     
where we have replaced $\textnormal{Im} \Sigma^R_a(\epsilon)$, which appears 
in the exponent after integration, by its zero-field value $-\hbar/2\tau_0$.
The oscillatory term in the above expression is slightly different from 
the usual expression, which is periodic in $\hbar \omega_c$, due to the 
presence of the Rashba and Zeeman splittings. 
In what follows we will ignore this modification of the density of states,
putting $\lambda=0$ and $\Delta_z=0$ in the oscillatory term.  
This does not mean, of course, that we are ignoring SO interactions 
altogether, since they still enter in the matrix elements of $\Gamma$.
Then the density of states becomes independent of the index $a$ and 
is given by the following simple expression:
\begin{equation}
\label{eq:22}
\varrho(\epsilon)=\varrho_0\left[1 + 2 \sum_{p=1}^{\infty} e^{-\pi p/\omega_c
 \tau_0} \cos\left(\frac{2\pi p \epsilon}{\hbar \omega_c} - \pi p
\right)\right],
\end{equation}
where $\varrho_1(\epsilon)=\varrho_2(\epsilon)=\varrho(\epsilon)/2$ and 
$\varrho_0=m/\hbar^2 \pi$ is the zero-field density of states.
If the magnetic field is not too large, i.e. if $\omega_c\tau \lesssim 1$, 
only the first term in the sum over $p$ in Eq.(\ref{eq:22}) needs to be 
retained. 
Then the oscillatory term in the density of states is purely sinusoidal.   
Substituting Eq.(\ref{eq:22}) in Eq.(\ref{eq:16}) we find that SCBA 
self-energy is also independent of the index $a$ and is given by:
\begin{equation}
\label{eq:23}
\textnormal{Im}\Sigma^R(\epsilon) \equiv - \frac{\hbar}{2\tau(\epsilon)} =
- \frac{\pi \hbar^3}{2 m \tau_0} \varrho(\epsilon).
\end{equation}  
The above disorder self-energy reduces to the usual expression $-\hbar/2\tau_0$
in the zero-field limit. At finite fields, the oscillatory term in the 
density of states in Eq.({\ref{eq:23}) is at the origin of a number of 
magnetoquantum oscillation phenomena, including the SdH oscillations of the 
electrical resistivity. 
As will be shown later, it also leads to similar magnetoquantum oscillation
effects in the spin relaxation rate.

\section{Spin density response function}
\label{sec:3} 
We can now calculate spin density response function of a disordered 2DEG system
with the Rashba SO interactions in a perpendicular magnetic field.
Transverse and longitudinal spin relaxation rates can be found from the 
poles of the corresponding spin density response functions. 
The calculation is similar to the analogous calculation that can be done at 
zero field. \cite{Burkov03}
We will use $\hbar=1$ units in the rest of this section, except for the final 
results.

We start from the general density matrix response function:
\begin{eqnarray}
\label{eq:24}
&&\chi_{\sigma_1\sigma_2,\sigma_3\sigma_4}({\bf r}-{\bf r}',t-t') = 
\nonumber \\ 
&-&i\theta(t-t') \langle \left[\hat \varrho^{\dag}_{\sigma_1
\sigma_2}({\bf r},t),\hat \varrho^{\vphantom \dag}_{\sigma_3\sigma_4}({\bf r}'
,t')\right]
\rangle,
\end{eqnarray}
where $\hat \varrho_{\sigma_1\sigma_2}({\bf r},t)=\Psi^{\dag}_{\sigma_2}
({\bf r},t) \Psi_{\sigma_1}({\bf r},t)$ is the generalized density operator,
whose expectation value is the density matrix.  
As usual, the calculation is most conveniently done by analytical 
continuation of the imaginary time response function. \cite{Senna82}
Fourier transformed imaginary time response function can be written as:
\begin{equation}
\label{eq:25}
\chi_{\sigma_1\sigma_2,\sigma_3\sigma_4}({\bf q},i\Omega) = 
\frac{1}{\beta} \sum_{i\omega} P_{\sigma_1\sigma_2,\sigma_3\sigma_4}
({\bf q},i\omega,i\omega+i\Omega),
\end{equation}
where 
\begin{eqnarray}
\label{eq:26}
&&P_{\sigma_1\sigma_2,\sigma_3\sigma_4}({\bf r}-{\bf r}',i\omega,i\omega+ 
i\Omega) = \nonumber \\
&&\langle {\cal G}_{\sigma_3\sigma_1}({\bf r}',{\bf r},i\omega)
{\cal G}_{\sigma_2\sigma_4}({\bf r},{\bf r}',i\omega+i\Omega)\rangle
\end{eqnarray} 
is the polarization bubble diagram  
and ${\cal G}$ are imaginary time Green's functions. 
$P$ can be calculated by summing all the SCBA diagrams 
for the disorder self-energy and all the ladder vertex corrections 
to the polarization bubble, which constitutes a conserving approximation 
for the density matrix response function.  
The result can be written in matrix notation as $P=P^0 D$, where 
$P^0$ is the ``bare'' polarization bubble with only the self-energy corrections
included:
\begin{eqnarray}
\label{eq:27}
&&P^0_{\sigma_1\sigma_2,\sigma_3\sigma_4}({\bf r}-{\bf r}',i\omega,i\omega+ 
i\Omega) = \nonumber \\
&&{\cal G}_{\sigma_3\sigma_1}({\bf r}',{\bf r},i\omega)
{\cal G}_{\sigma_2\sigma_4}({\bf r},{\bf r}',i\omega+i\Omega),
\end{eqnarray} 
with ${\cal G}$ here being the disorder-averaged Green's function.
The vertex part, or, as it is often called, diffusion propagator, $D$ 
satisfies the following matrix equation:
\begin{equation}
\label{eq:28}
D = {\bf 1} + \gamma P^0 D,
\end{equation}
where $\gamma \equiv \langle V^2({\bf r})\rangle$.
The solution of this equation is 
\begin{equation}
\label{eq:29}
D = \left[{\bf 1} - \gamma P^0\right]^{-1}.
\end{equation}

\begin{figure}[t]
\includegraphics[width=8cm]{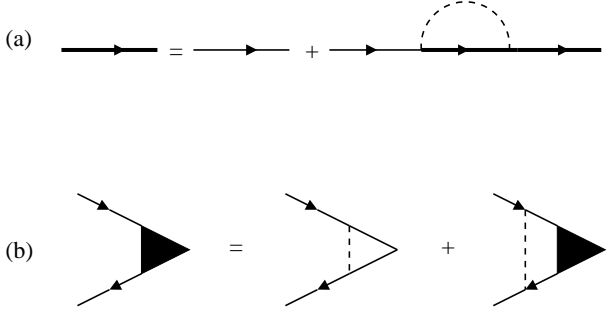}
\caption{(a) SCBA equation for the disorder-averaged Green's function; 
(b) Equation for the vertex function $D$. Dashed lines denote impurity 
potential correlator $\langle V({\bf r}) V({\bf r}')\rangle$.} 
\label{diagram}
\end{figure}

We can now calculate the retarded real time response function by analytically 
continuing Eq.(\ref{eq:25}) to real frequencies. 
The result can be written as an integral along the branch cut of the 
SCBA Green's function:
\begin{eqnarray}
\label{eq:30}
&&\chi({\bf q},\Omega) = \int_{-\infty}^{\infty} \frac{d \epsilon}{2 \pi i}
n_F(\epsilon) \left[P({\bf q},\epsilon+i\eta,\epsilon+\Omega+i\eta)\right. 
\nonumber \\
&-&\left.P({\bf q},\epsilon-i\eta,\epsilon+\Omega+i\eta)
+P({\bf q},\epsilon-\Omega-i\eta,\epsilon+i\eta)\right. \nonumber \\ 
&-&\left.P({\bf q},\epsilon-\Omega-i\eta,\epsilon-i\eta)\right].
\end{eqnarray}   
At low frequencies and low temperatures Eq.(\ref{eq:30}) can be simplified to:
\begin{equation}
\label{eq:31}
\chi({\bf q},\Omega) = \frac{i\Omega}{2\pi}P({\bf q},\epsilon_F-i\eta,
\epsilon_F+\Omega+i\eta)+\frac{\varrho(\epsilon_F)}{2}.
\end{equation} 
Thus, the problem reduces to calculating the matrix 
$P^0({\bf q},\epsilon_F-i\eta,\epsilon_F+\Omega+i\eta)$.
We will in fact be interested only in the relaxation of uniform spin 
polarization and thus will put ${\bf q} = 0$ henceforth. 
The calculation of $P^0$ in our case closely resembles
the analogous calculation in the zero-field problem. \cite{Burkov03}

We start from the definition of $P^0$ Eq.(\ref{eq:27}), using 
the disorder-averaged Green's functions calculated in section \ref{sec:2}:
\begin{eqnarray}
\label{eq:32}
&&P^0_{\sigma_1\sigma_2,\sigma_3\sigma_4}(\epsilon_F-i\eta,
\epsilon_F+\Omega++i\eta)
\nonumber \\
&=&\frac{1}{L^2} \int d{\bf r} d {\bf r}' G^A_{\sigma_3\sigma_1}({\bf r}',
{\bf r},\epsilon_F) G^R_{\sigma_2\sigma_4}({\bf r},{\bf r}',\epsilon_F+\Omega)
\nonumber \\
&=&\frac{1}{L^2} \int d{\bf r} d{\bf r}' \sum_{nak,n'a'k'} 
\frac{\Psi^{\vphantom *}_{nak}({\bf r}',\sigma_3) 
\Psi^*_{nak}({\bf r},\sigma_1)}
{\epsilon_F -\epsilon_{na} - i/2\tau} \nonumber \\
&\times&\frac{\Psi^{\vphantom *}_{n'a'k'}({\bf r},\sigma_2) 
\Psi^*_{n'a'k'}({\bf r}',
\sigma_4)}{\epsilon_F +\Omega-\epsilon_{n'a'} + i/2\tau},
\end{eqnarray}
where $\tau \equiv \tau(\epsilon_F)$ and the wavefunctions are given by 
Eq.(\ref{eq:7}). 
The spatial integrals in the above equation can be easily evaluated 
using the orthonormality of the Landau gauge eigenfuctions and 
are given by:
\begin{eqnarray}
\label{eq:33}
\int d{\bf r} \Psi^*_{nak}({\bf r},\uparrow) 
\Psi^{\vphantom *}_{n'a'k'}({\bf r},\uparrow)
&=& v^*_a v^{\vphantom *}_{a'} \delta_{n,n'} \delta_{k,k'},\nonumber \\
\int d{\bf r} \Psi^*_{nak}({\bf r},\downarrow) 
\Psi^{\vphantom *}_{n'a'k'}({\bf r},
\downarrow)&=& u^*_a u^{\vphantom *}_{a'} \delta_{n,n'} \delta_{k,k'},
\nonumber \\
\int d{\bf r} \Psi^*_{nak}({\bf r},\uparrow) 
\Psi^{\vphantom *}_{n'a'k'}({\bf r},
\downarrow)&=& v^*_a u^{\vphantom *}_{a'} \delta_{n,n'+1} 
\delta_{k,k'},\nonumber \\
\int d{\bf r} \Psi^*_{nak}({\bf r},\downarrow) 
\Psi^{\vphantom *}_{n'a'k'}({\bf r},
\uparrow)&=& u^*_a v^{\vphantom *}_{a'} \delta_{n,n'-1} \delta_{k,k'}.
\end{eqnarray}
The sums over Landau level indices in Eq.(\ref{eq:32}) can be done as follows. 
We first decouple the product of two Green's functions appearing in 
(\ref{eq:32}) as:
\begin{eqnarray}
\label{eq:34}
&&\frac{1}{(\epsilon_F-\epsilon_{na}-i/2\tau)(\epsilon_F+\Omega-
\epsilon_{n'a'}+i/2\tau)} \nonumber \\
&=&\frac{1}{\Omega + (-1)^a\Delta - (-1)^{a'}\Delta + i/\tau} \nonumber \\ 
&\times&\left(
\frac{1}{\epsilon_F-\epsilon_{na}-i/2\tau} - \frac{1}{\epsilon_F+\Omega-
\epsilon_{n'a'}+i/2\tau} \right). \nonumber \\
\end{eqnarray}
The real parts of the decoupled sums over Landau level indices can then be 
neglected by our assumption that 
$\epsilon_F \gg \hbar \omega_c, \lambda p_F, \hbar \Omega$.
The imaginary parts are simply proportional to the density of 
states at Fermi energy $\varrho(\epsilon_F)$.
Thus we obtain the following expressions for the matrix elements 
of $P^0$:
\begin{eqnarray}
\label{eq:35}
\gamma P^0_{\uparrow\uparrow,\uparrow\uparrow} &=& \gamma
P^0_{\downarrow\downarrow,\downarrow\downarrow} = \left(1-\frac{1}{2}
\sin^2(2\vartheta)\right)f_0(\Omega) \nonumber \\ 
&+&\frac{\sin^2(2 \vartheta)}{4} 
\left[f_+(\Omega) + f_-(\Omega)\right],\nonumber \\
\gamma P^0_{\uparrow\uparrow,\downarrow\downarrow} &=& 
\gamma P^0_{\downarrow\downarrow,\uparrow\uparrow} = 
\frac{\sin^2(2\vartheta)}{4} \left[2 f_0(\Omega) - f_+(\Omega) - 
f_-(\Omega)\right], \nonumber \\
\gamma P^0_{\uparrow\downarrow,\uparrow\downarrow} &=& 
\frac{\sin^2(2\vartheta)}{2} f_0(\Omega+\omega_c) + \sin^4(\vartheta) 
f_+(\Omega+\omega_c) \nonumber \\
&+& \cos^4(\vartheta) f_-(\Omega+\omega_c),\nonumber \\
\gamma P^0_{\downarrow\uparrow,\downarrow\uparrow} &=& 
\frac{\sin^2(2\vartheta)}{2} f_0(\Omega-\omega_c) + \sin^4(\vartheta) 
f_-(\Omega-\omega_c) \nonumber \\
&+& \cos^4(\vartheta) f_+(\Omega-\omega_c),
\end{eqnarray} 
where functions $f_0$ and $f_{\pm}$ are given by:
\begin{eqnarray}
\label{eq:36}
f_0(\Omega) &=& \frac{1}{1-i\Omega \tau}, \nonumber \\
f_{\pm}(\Omega) &=& \frac{1}{1 - i\Omega \tau \pm i\Delta \tau}.
\end{eqnarray}
All other matrix elements of $P^0$ vanish.

To calculate the transverse and longitudinal spin relaxation rates we need to 
find the poles of the 
spin density response function, or, equivalently, zeroth of the inverse 
diffusion propagator $D^{-1}$. 
It is most easily done by transforming $D^{-1}$ to the physical space 
of charge and spin density components, which is accomplished by multiplying 
it on both sides with Pauli matrices:
\begin{equation}
\label{eq:37}
D^{-1}_{\alpha\beta} = \frac{1}{2} \tau^{\alpha}_{\sigma_1\sigma_2} 
D^{-1}_{\sigma_1\sigma_2,\sigma_3\sigma_4} \tau^{\beta}_{\sigma_4\sigma_3},
\end{equation}
where $\alpha,\beta=c,x,y,z$ and $\tau^c$ is the identity matrix.
Thus we obtain:
\begin{eqnarray}
\label{eq:38}
D^{-1}_{cc}(\Omega)&=&1 - f_0(\Omega), \nonumber \\
D^{-1}_{zz}(\Omega)&=&1 - f_0(\Omega) + \frac{\sin^2(2\vartheta)}{2} 
\left[2 f_0(\Omega)\right. \nonumber \\ 
&-&\left.f_+(\Omega) - f_-(\Omega)\right], \nonumber \\
D^{-1}_{+-}(\Omega)&=&1 - \frac{\sin^2(2\vartheta)}{2} f_0(\Omega+\omega_c)-
\sin^4(\vartheta) f_+(\Omega+\omega_c) \nonumber \\
&-&\cos^4(\vartheta) f_-(\Omega+\omega_c).
\end{eqnarray}
Clearly $D^{-1}_{cc}(0) = 0$, which is a consequence of charge conservation. 
For the spin part of the inverse diffusion propagator, to leading order in 
$\lambda p_F \tau /\hbar$ and $\Omega \tau$, and assuming that 
$\Delta_z \ll \hbar \omega_c$, we obtain:
\begin{eqnarray} 
\label{eq:39}
\tau^{-1} D^{-1}_{zz}(\Omega) &=& -i\Omega + \frac{1}{\tau_z}, \nonumber \\
\tau^{-1} D^{-1}_{+-}(\Omega) &=& -i\Omega + i\hbar^{-1} \Delta_z +
\frac{1}{\tau_{\perp}},
\end{eqnarray}
where the longitudinal and transverse spin relaxation rates are given by:
\begin{eqnarray}
\label{eq:40}
\frac{1}{\tau_z} &=& \frac{8 \lambda^2 m \epsilon_F \tau /\hbar^2}
{1 + (\omega_c \tau)^2}, \nonumber \\
\frac{1}{\tau_{\perp}} &=& \frac{1}{2\tau_z}.
\end{eqnarray}
Eq.(\ref{eq:40}) is our main result.
The form of these expressions is rather similar to the well-known 
Dyakonov-Kachorovskii expressions \cite{Dyakonov86} for the spin relaxation 
rates in the zero field case, except for the factor $1/[1+(\omega_c \tau)^2]$,
which describes the suppression of spin relaxation by the perpendicular 
field. 
The mechanism of this suppression can be understood in quasiclassical 
terms by imagining electrons moving along classical cyclotron orbits 
and undergoing impurity scattering and the Rashba spin precession.
Orbital motion affects the spin precession, since the direction of
the Rashba field is always transverse to the direction of electron's velocity,
and thus changes as the electron moves along a circular cyclotron orbit.
However, if $\omega_c \tau \ll 1$, impurity scattering randomizes the 
direction of electron's motion and the influence of the magnetic field
on the orbital motion is then negligible. 
On the other hand, when $\omega_c\tau \gg 1$, electrons can move around 
cyclotron orbits almost freely and thus the Rashba spin precession is 
averaged out.
This picture of the suppression of the DP spin relaxation by a perpendicular 
magnetic field has been discussed before by several authors. \cite{Ivchenko73,Margulis83,Martin02}
However, we are not aware of a rigorous derivation of a simple analytical
expression for the spin relaxation rate, such as Eq.(\ref{eq:40}).
     
Probably the most noteworthy feature of our result is the oscillatory 
dependence on the magnetic field, which appears due to the oscillatory 
term in the scattering rate $1/\tau$ in Eqs.(\ref{eq:22}) and (\ref{eq:23}).
If $\omega_c \tau < 1$, we may neglect the $(\omega_c\tau)^2$ term in the 
denominator in Eq.(\ref{eq:40}) and leave only the first term in the sum over 
$p$ in Eq.(\ref{eq:22}). 
In this case spin relaxation rates will exhibit purely sinusoidal SdH-like
oscillations.
Similar effects have been discussed earlier in the context 
of nuclear spin-lattice relaxation in quantum Hall systems.\cite{Vagner88,Klitzing90,MacDonald91} 
However, the possibility of such magnetoquantum oscillations in the SO-induced
spin relaxation rates has never been discussed before. 
Such oscillations have in fact been observed in recent experiments 
on spin relaxation in InGaAs quantum wells.\cite{Awschalom03} 

In conclusion, we have considered the problem of spin relaxation in 
a 2DEG with the Rashba SO interactions in a perpendicular magnetic field. 
We have given a rigorous microscopic derivation of the transverse and 
longitudinal spin relaxation rates from the poles of the spin density 
response function in the limit $\epsilon_F \gg \hbar \omega_c$. 
It is possible to extend our calculation to other regimes, 
including the quantum Hall regime $\epsilon_F \sim \hbar \omega_c$.
Our results demonstrate that magnetoquantum oscillation effects should 
be observed in spin relaxation phenomena in 2DEGs.
One of the advantages of our approach is that it can be easily extended to 
study another very interesting and still unexplored question of what is the 
effect of Coulomb interactions on the spin relaxation in semiconductor 2DEGs. 
This question will be addressed in a future publication. 
      
We would like to acknowledge helpful discussions with David Awschalom, 
Yogesh Joglekar, Allan MacDonald and Vanessa Sih.
This work was supported by DARPA/ONR N00014-99-1-1096, by the NSF under 
grant DMR-9985255 and by the Sloan and Packard foundations.

\end{document}